\documentclass{elsart}

\usepackage{harvard}

\usepackage{graphicx}

\usepackage{amssymb}

\def\url#1{{\ttfamily\def\/{/\discretionary{}{}{}}#1}}

\begin{document}

\begin{frontmatter}
\title{X-rays from Radio-Galaxies: BeppoSAX Observations}


\author[paola]{P. Grandi\thanksref{pg}}, 
\author[meg]{C. M. Urry},
\author[laura]{L. Maraschi}
\thanks[pg]{E-mail: grandi@alphasax2.ias.rm.cnr.it, paola@ias.rm.cnr.it}

\address[paola]{IAS/CNR, Area di Ricerca Tor Vergata, I-00133, Roma , Italy}
\address[meg]{Space Telescope Science Institute, 3700 San Martin Dr., 
Baltimore, MD 21218}
\address[laura]{Osservatorio Astronomico di Brera, via Brera 28, I-20121, Milano, Italy}

\begin{abstract}
We briefly review BeppoSAX observations of X-ray bright radio-galaxies.
Their X-ray spectra are quite varied, and perhaps surprisingly,
any similarity between radio-loud AGN and Seyfert galaxies is 
the exception rather than the rule.
When detected, reprocessing features (iron line and reflection) are  
generally weak, suggesting two possible scenarios:
either: (1) non-thermal (jet?) radiation dilutes the X-ray emission 
from the disk in radio-loud objects, or (2) the solid angle subtended by 
the X-ray reprocessing material is smaller in radio-loud than in radio-quiet 
AGN due to different characteristics of the accretion disk itself.

\end{abstract}

\end{frontmatter}

\section{Introduction}
\label{intro}

X-ray observations of radio-galaxies can play an important role in
understanding the AGN radio-loud and radio-quiet dichotomy.
Several authors have speculated upon the possibility that
the AGN distinction in two big families
reflects different physical properties of the nuclear
engine. Several years ago, Rees et al. (1982),
suggested that in the nuclei of radio-galaxies
a spinning black hole is surrounded by a hot thick accretion flow,
in contrast with the radio-quiet picture, which
assumes the accreting material in shape of cold thin disk.
Later, other authors (Blandford 1990, Meier these proceedings)
indicated the black hole spin as the physical parameter 
responsible of the AGN dichotomy, being more rapid in radio-louds 
and therefore more efficient in producing powerful jets.

X-ray photons are produced in the inner regions
($<1$ pc) of AGN and provide crucial information on
physical processes occurring very near to the black hole.
X-ray studies of radio-galaxies are then essential
to answer a fundamental question:
are the same accretion processes at work in radio-loud
and radio-quiet AGNs?

In this paper, we present an on-going analysis
of radio-galaxies observations performed with 
the BeppoSAX satellite (0.1-150 keV) and discuss
the results of a comparative study between our sources
and a sample of 12 Seyfert galaxies \cite{mat99} observed with the same
satellite.

\section{Results}
\label{structure}

Our sample (Table 1) consists of 6 Broad Line Radio Galaxies (BLRG),
the radio-loud counterpart of Seyferts 1 in the AGN Unified Schemes.
Most of the sources have a Fanaroff-Riley (FR) II radio morphology
and 3 shows superluminal motions.
For comparison, one Narrow Line Radio Galaxy (NLRG), Centaurus A (Cen A),
is also included in the sample.
As two observations are available for this source,
a spectral variability study was also possible.

\begin{table*}
\begin{center}
\caption{}
\begin{tabular}{lcccc} 
\hline 

Source & z & Optical   & Radio$^a$     & L$^b_{2-10 keV}$  \\
       &   & Type      & Morphology    & erg cm$^{-2}$ sec$^{-1}$)\\
&&&\\
PKS2152-69       & 0.027 & BLRG & FRI/FRII  & $0.2\times10^{44}$   \\
Pictor A         &0.035  & BLRG & FRII  & $1.0\times10^{44}$   \\
3C120            &0.033  & BLRG & FRI/S & $2.4\times10^{44}$  \\
3C111            &0.048  & BLRG & FRII/S& $2.9\times10^{44}$   \\ 
3C390.3          &0.057  & BLRG & FRII/S& $3.3\times10^{44}$   \\
3C382            & 0.059 & BLRG & FRII  & $9.4\times10^{44}$    \\
Cen A            & 0.002& NLRG & FRI   & $0.5\times10^{43}$    \\
\hline

\multicolumn{5}{l}{$^a$ -- S = superluminal source; $^b$ -- 
Luminosity corrected for absorption}\\
\end{tabular}
\end{center}
\end{table*}

The X-ray spectra of radio-galaxies are complex (Fig. 1)
and varied, as shown in Table 2 where the spectral 
fit results are listed.
\begin{figure}
\begin{center}
\includegraphics*[width=8cm,angle=-90]{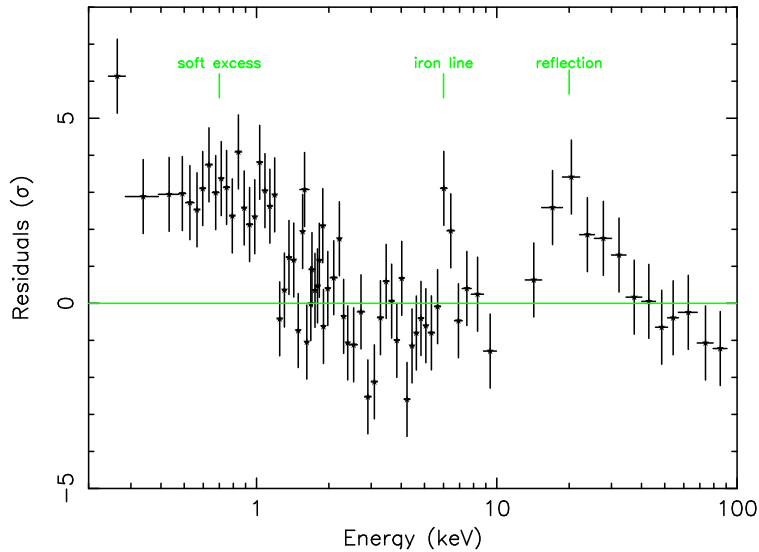}
\end{center}
\caption{The complex spectrum of 3C382.
When the BeppoSAX data are  fitted
with a simple power law, residuals show an excess of emission
at soft energies, an iron line
at $\sim 6.4~$ kev and a reflection hump above 10 keV}
\end{figure}
A simple power law  plus Galactic absorption generally
did not give good fits to the data. 
We then tested a more complex model which takes into account all the
spectral components usually observed in Seyfert 1 galaxies:
a power law which an exponential cutoff (Cutoff), a
fluorescence K$_{\alpha}$ iron line and a reflection component (Refl).
The theoretical picture which interprets the Seyfert spectra assumes 
a cold thin optically thick disk with a hot corona above it.
The corona produces the X-ray photons by inverse Compton of the UV disk 
radiation. The disk itself reprocesses the 
down-scattered X-ray radiation 
producing the iron line and a hump above $\sim 10$ keV \cite{ha91,ha93}.

A multicomponent spectrum is generally a good representation of the data,
although all the spectral features were not always simultaneously present 
in each source.
In some cases, it was also necessary to include a soft excess
(parameterized with a steep power law $\Gamma_{\it soft}=2-3$)
and/or to include an extra absorption in addition to the Galactic one (see 
Table 2).
\begin{table*}
\begin{center}
\footnotesize
\caption{}
\begin{tabular}{lcccccccc} 
\hline
Source & $\Gamma$ & N$_H$ & Refl. & Cutoff
& E$_{Fe}$ & $\sigma$ &  EW \\
          &          & ($10^{21}$ cm$^{-2}$) & &  keV&
keV & keV & eV \\
&&&&&&&\\
PKS2152-69$^{c}$ &  1.79$\pm0.1$& 0.25 (f) &... &... &
6.4 (f) & 0.0 (f) & $<429$ \\
Pictor A & 1.63$\pm0.06$& 0.42 (f)&...  & ...  &
6.4 (f) & 0 (f) & $<102$\\
3C120$^{\star}$ & 1.80$^{+0.12}_{-0.30}$ & 2.4$^{+1.8}_{-1.3}$ & 0.7$\pm0
.4$ & 109$^{+125}_{-77}$ & 
6.2$\pm0.4$ & 0.27$^{+0.40}_{-0.27}$ &59$^{+82}_{-42}$\\
3C111$^{\dag}$& 1.65$\pm0.04$ &7.1$^{+0.9}_{-0.8}$ &$<0.3$ & $>90$ &
6.6$^{+0.4}_{-0.2}$ & 0 (fixed) & 58$^{+31}_{-55}$\\
3C390.3& 1.80$^{+0.05}_{-0.04}$ & 1.3$\pm0.2$ & 1.2$^{+0.4}_{-0.3}$& $>12
3$&6.4$\pm0.1$ & 0.07$^{+0.21}_{-0.07}$& 136$^{+40}_{-36}$\\
3C382$^{\star}$& 1.79$\pm0.04$& 0.88 (f) & 0.4$^{+0.3}_{-0.2}$&
155$^{+148}_{-59}$&
6.5$^{+0.9}_{-0.2}$& 0.00 (f) &31$^{+44}_{-15}$  \\
Cen A(1997) &  1.76$^{+0.04}_{-0.06}$ & 98$^{+20}_{-30}$ & 0.2$\pm0.1$ &
286$^{+257}_{-94}$ &
6.48$\pm0.07$ & 0.3$\pm0.01$ & 166$^{+38}_{-32}$\\
Cen A(1998)  & 1.73$^{+0.05}_{-0.04}$  & 92$^{+30}_{-10}$&
0.1$\pm0.1$  & 297$^{+156}_{-79}$ &
6.38$\pm0.08$ &0.08$^{+0.26}_{-0.08}$ &  $57\pm18$\\\hline
\multicolumn{9}{l}{$^a$ -- photons cm$^{-2}$ sec$^{-1}$ keV$^{-1}$;
$^b$ -- ergs cm$^{-2}$ sec$^{-1}$ $\AA^{-1}$; $^c$ -- MECS data only}\\
\multicolumn{9}{l}{$^\star$ -- Sources with detected soft excess; 
$^\dag$ -- iron line marginal detection}\\
\end{tabular}
\end{center}
\end{table*}
\normalsize

{\it Continuum of emission} -- The shape of the BLRG primary X-ray 
continuum does not significatively differ from that of Seyfert 
1 galaxies \cite{mat99}.
The BLRG spectral slopes and the cutoff energies (when detected) are 
consistent with those observed in Seyferts 1 ($<\Gamma^{\it Sey}=1.85>$, 
$\sigma^{\Gamma}_{rms}=0.22$; $<E^{\it Sey}_{\it cutoff}=237>$ keV,
$\sigma^{\it cutoff}_{rms}=150$). 

{\it Soft excess} -- Soft excesses were found in 3C120 and 3C382.
Since these are the AGNs in our sample with strong UV bumps \cite{ma91,tad86},
we suggest that the soft photons represent 
the hard tail of the thermal emission associated with the accretion disk.

{\it Cold Absorber} -- A cold absorbing column in excess of Galactic is
observed not only in the NLRG Centaurus A, but also in half the BRLGs 
of the sample. In contrast, we did not find any signatures of warm 
absorber which is rather common in Seyferts \cite{mat99}.
This result suggests that the nuclear environments of radio-loud 
and radio-quiet AGN might be different.


{\it Iron Line and Reflection Hump} -- 
While in Seyfert 1 sample the iron line is always detected, 
in BLRGs the reprocessed feature is detected in half the sources. 
In addition, in BLRGs the iron line equivalent widths are significantly 
smaller than in Seyfert 1s
($<EW^{\it BLRG}=71>$ eV, $\sigma^{\it BLRG}_{rms}=45$;
$<EW^{\it Sey1}=175>$ eV, $\sigma^{\it Sey1}_{rms}=52$).
The weakness of the iron line in BLRGs indicates that a simple
Seyfert-like picture is not suitable to describe
the nuclear region of radio-galaxies.

All the BLRGs with detected iron lines show a 
reflection  bump. Although the strength of the reflection is not
very well constrained, there is indication that weak reflection 
corresponds to weak iron line (see Table 2).
This is consistent with the Fe line arising in the same 
material that reflects (and reprocesses) the primary X-ray continuum.

{\it Iron Line Variability} -- Iron line variability studies
are extremely powerful to investigate the location of the emitting gas.
In the BeppoSAX observations of Cen A (Table 2)
the iron line was more intense when the source was weaker.
The total number of  photons in the line was about twice as large 
as in the first BeppoSAX observation when the nuclear flux of 
the source was 25$\%$ lower (see fig.1 in Grandi et al. 1998).
{\bf This lack of direct correlation shows that the iron line emitting region 
responds with a significant delay to the continuum variations and 
therefore cannot be located near the primary X-ray source.}
A similar conclusion was reached by Wozniak et al. (1998) 
for 3C390.3, for which ASCA and Ginga observations
did not reveal any long term variability in the Fe line when the 
continuum flux changed.

The comparison of our data with previous observations by GINGA \cite{nan94} 
and ASCA  \cite{gra97} shows that 
a  similar lack of correlation between the continuum flux and 
the Fe equivalent is present also in 3C120 and 3C382.
In both radio-galaxies, BeppoSAX measured a continuum flux higher than ASCA and GINGA, but significantly weaker Fe equivalent width.
3C120 and 3C382 have spectral characteristics of Seyfert 1s.
Both have a huge UV bump and a soft excess probably produced by a 
cold accretion disk. In addition, ASCA detected a broad iron line in both 
these sources \cite{gra97,re98}, as expected 
if produced in the inner region of the accretion flow.

In this case, {\bf the observed dilution of the equivalent 
width could be  produced by an increase of the jet intensity rather 
than by a delayed response  of the cold matter to variations 
of the X-ray source}.


\section{Discussion}

BeppoSAX analysis of radio-galaxies has 
pointed out important differences between radio-loud  and radio-quiet AGN
and has clearly shown, for the first time,  that
in BLRGs the iron line equivalent widths (and probably the reflection 
strengths) are significantly smaller than in Seyferts 1. 

In some radio-galaxies the cold material  responsible for the line 
production is located far from the X-ray primary source.
It is possible that the innermost part of the accretion flow is not
in shape of cold disk. 
The accretion flow might be then  hot and geometrically thick in the 
inner regions \cite{rees82,sha76,na98} and 
cold optically thick \cite{shak73} only at larger radii, subtending
relatively smaller solid angles (compared to Seyfert 1s).
Featureless spectra or  weak iron lines could be signatures of low efficiency
accretion processes.

In sources with Seyfert-like spectra (i.e, 
with soft excess, strong UV bump and broad iron lines), the 
Fe line might be produced by a cold thin optically thick disk (Seyfert-like
accretion flow).
In this case, the iron line EW might be diluted by the
Doppler-enhanced non-thermal jet continuum.
This explanation predicts that the observed continuum (disk +jet) is more 
variable than the Fe line (disk).

\end{document}